\documentclass[conference]{IEEEtran}
\IEEEoverridecommandlockouts
\usepackage{cite}
\usepackage{amsmath,amssymb,amsfonts}
\usepackage{mathrsfs}
\usepackage{algorithm}
\usepackage{graphicx}
\usepackage{algpseudocode}
\usepackage{textcomp}
\usepackage{xcolor}
\def\BibTeX{{\rm B\kern-.05em{\sc i\kern-.025em b}\kern-.08em
    T\kern-.1667em\lower.7ex\hbox{E}\kern-.125emX}}
\newtheorem{theorem}{Theorem}
\newtheorem{lemma}{Lemma}

\begin{document}
\title{
Statistical Design of Multi-BD-RIS-Assisted User-Centric Cell-Free Massive MIMO
}
\author{\IEEEauthorblockN{Mohammad Ezaz, Azadeh Vosoughi}
\IEEEauthorblockA{\textit{Department of Electrical \& Computer Engineering, University of Central Florida} \\
smezaz@ucf.edu, azadeh@ucf.edu}
\thanks{This work has been submitted to the IEEE for possible publication.
Copyright may be transferred without notice, after which this version may
no longer be accessible.}
}
\maketitle
\begin{abstract}
We study the uplink of a user-centric (UC) cell-free massive MIMO system assisted by multiple beyond-diagonal reconfigurable intelligent surfaces (BD-RISs), under imperfect channel state information (CSI) and pilot reuse. The BD-RISs are modeled as reciprocal, lossless, group-connected scattering networks. We first derive the minimum mean-square error (MMSE) estimation statistics for each user's aggregate channel to its serving access points (APs). To avoid frequent CSI acquisition and RIS reconfiguration, we optimize the BD-RIS scattering matrices using long-term channel statistics, allowing their configurations to remain fixed over multiple coherence blocks. Specifically, we minimize the sum of normalized mean-square errors (NMSEs) of the MMSE  aggregate channel estimators, subject to the symmetry and unitarity constraints of the group-connected architecture. We develop a product-manifold Riemannian optimization algorithm and analyze its convergence and computational complexity. To evaluate uplink spectral efficiency (SE), we separately derive a closed-form expression for the effective SINR under maximum ratio (MR) combining at the central processing unit (CPU) and establish a lower bound on the achievable ergodic SE. The SINR derivation accounts for the statistical dependence among channels associated with different APs due to shared BD-RIS-assisted propagation paths, including cross-AP second-order statistics under pilot reuse. Numerical results show that moderate group connectivity retains most of the NMSE improvements achieved by full connectivity, with larger optimization gains as BD-RIS-assisted paths strengthen relative to direct AP–user links. They also reveal that lower estimation NMSE does not necessarily lead to higher uplink SE.
\end{abstract}
\section{Introduction}
Reconfigurable intelligent surface (RIS)-assisted user-centric (UC) cell-free massive MIMO has been widely studied for channel estimation, performance analysis, and statistical resource allocation under imperfect channel state information (CSI). RIS-assisted cell-free systems with spatially correlated channels were analyzed in~\cite{chien2021riscf}, while~\cite{dai2023two} developed a two-timescale transmission design based on long-term channel statistics. Spatially correlated Rician fading was considered in~\cite{cha2026}, and~\cite{alnahhas2024} investigated multi-RIS deployments in cell-free networks. More recently,~\cite{tran2026ul-ris-cf} jointly optimized UC access point (AP) clustering, large-scale fading decoding, conventional RIS phase shifts, and uplink power optimization using statistical CSI. These works demonstrate the potential of RIS-assisted cell-free systems, but rely primarily on conventional RISs.

Beyond-diagonal (BD)-RIS architectures provide greater propagation flexibility through inter-element coupling, allowing connected elements to jointly manipulate the incident waves rather than applying independent phase shifts as in conventional RISs. Their surface response is therefore characterized by a {\it scattering matrix} that generalizes the conventional diagonal phase-shift matrix~\cite{liBDRIS2022}. Single-, group-, and fully connected BD-RIS architectures were studied for rate, power, and energy-efficiency optimization in~\cite{zhou2023optimizing}. For reciprocal and lossless BD-RISs,~\cite{santamaria2026manifold} characterized the symmetric unitary manifold and developed a Riemannian optimization framework for a fully connected BD-RIS-assisted MIMO link, while~\cite{fidanovski2026} extended this framework to group-connected BD-RISs through a product of symmetric unitary block manifolds. These developments, however, have largely remained separate from  UC cell-free analysis under imperfect CSI.
In particular, existing RIS-assisted cell-free designs employ conventional RISs, whereas existing group-connected BD-RIS manifold formulations focus on cellular MIMO systems. 

To address this knowledge gap, we investigate a multi-BD-RIS-assisted UC cell-free uplink system. Since optimizing the BD-RIS scattering matrices using instantaneous small-scale CSI would require frequent channel acquisition and RIS reconfiguration, we instead rely on long-term channel statistics, allowing the configurations to remain fixed over multiple coherence blocks.
Specifically, we derive the minimum mean-square error (MMSE) estimation statistics for each user's aggregate channel to its serving APs. We then formulate an optimization problem that minimizes the sum of the normalized mean-square errors (NMSEs) of the MMSE  aggregate channel estimates, subject to the symmetry and unitarity constraints of the group-connected BD-RIS architecture. We derive the corresponding Euclidean and Riemannian gradients and develop a product-manifold optimization algorithm, together with convergence and complexity analyses.
Furthermore, for maximum ratio (MR) combining performed at the CPU, we derive a closed-form expression for the effective uplink SINR and establish a lower bound on the achievable ergodic spectral efficiency (SE). This derivation accounts for the statistical dependence among channels associated with different APs due to shared BD-RIS-assisted propagation paths, including cross-AP second-order statistics under pilot reuse.
Our numerical results demonstrate that the group-connected BD-RIS architecture retains most of the NMSE improvement achieved by full connectivity, with larger optimization gains as BD-RIS-assisted paths become stronger relative to direct AP–user links. They also reveal that lower NMSE does not necessarily translate into higher uplink SE.
%
%
%
\section{System Model}
\label{sec:system-model}
\begin{figure}[tbhp]
\centering
\vspace{-4mm}
\includegraphics[height=1.5in, keepaspectratio]{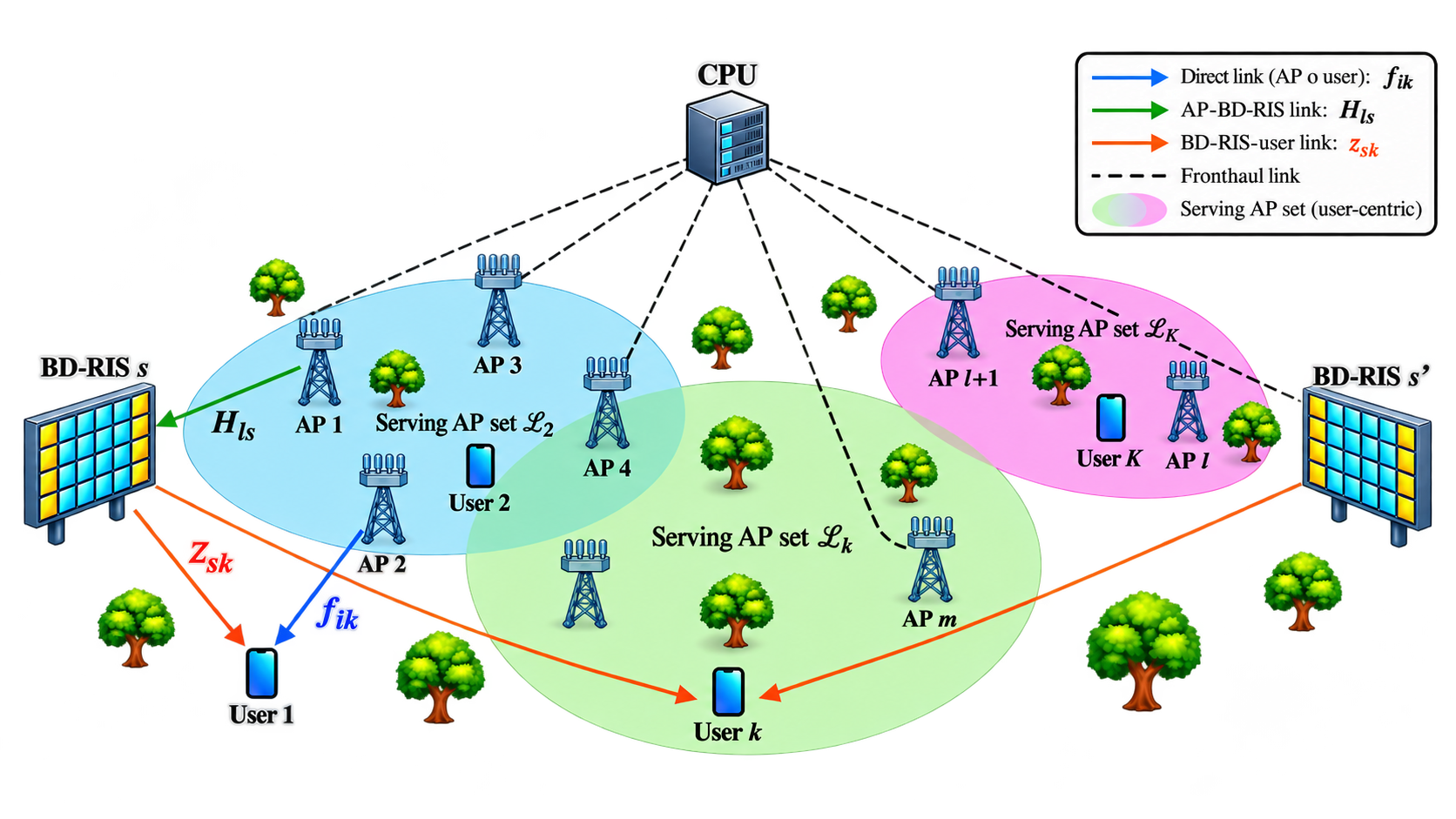}
\vspace{-5mm}
\caption{Multi-BD-RIS-assisted UC cell-free massive MIMO system.}
\label{fig:sysmod}
\vspace{-3mm}
\end{figure}

We consider the uplink of a multi-BD-RIS-assisted UC cell-free massive MIMO
system with $L$ distributed $M$-antenna APs, $K$ single-antenna users,
and $S$ BD-RISs, where $K\ll LM$. The APs are connected to a CPU through error-free fronthaul links. Let
$\mathscr L_k\subseteq\{1,\ldots,L\}$ denote the APs serving user $k$,
and $\mathscr K_l\triangleq\{k:l\in\mathscr L_k\}$ the users served by
AP $l$. The BD-RISs are fixed, while the APs and users are distributed
over the service area, as illustrated in Fig.~\ref{fig:sysmod}.
Each BD-RIS has $N$ elements and scattering matrix
$\boldsymbol{\Theta}_s\in\mathbb C^{N\times N}$. We adopt the
group-connected architecture~\cite{liBDRIS2022,shenBDRIS2021}, where
the elements are partitioned into $G$ groups of size
$\bar N=N/G$, such that
\(
\boldsymbol{\Theta}_s=
\operatorname{blkdiag}
(\boldsymbol{\Theta}_{s,1},\ldots,\boldsymbol{\Theta}_{s,G})
\)
with
$\boldsymbol{\Theta}_{s,g}\in\mathbb C^{\bar N\times\bar N}$.
For passive, reciprocal, and lossless BD-RISs, each block satisfies
\(
\boldsymbol{\Theta}_{s,g}
=
\boldsymbol{\Theta}_{s,g}^{T},
\
\boldsymbol{\Theta}_{s,g}^{H}\boldsymbol{\Theta}_{s,g}
=
\mathbf I_{\bar N},
\ \forall s,g,
\)
corresponding to symmetry from reciprocity and unitarity from
losslessness. The BD-RIS configuration is common to all users and fixed
within each coherence block.
We assume block fading with coherence length $\tau_c=\tau_p+\tau_u$,
where $\tau_p$ and $\tau_u$ are allocated to uplink training and data
transmission, respectively. The $\tau_p$ orthogonal pilots are reused
according to a predetermined pilot assignment, as detailed in
Section~\ref{subsec:uplink-training-channel-estimation}.
%
%
\subsection{Uplink Aggregate Channel Model}
\label{subsec:channel-model}
Let $\mathbf f_{lk}\in\mathbb C^{M}$ and $\mathbf z_{sk}\in\mathbb C^{N}$ denote the user $k$-AP $l$ and user $k$-RIS $s$ uplink channels, respectively. The AP $l$-RIS $s$ channel is $\mathbf H_{ls}\in\mathbb C^{N\times M}$, and  the reciprocal RIS $s$-AP $l$ channel is $\mathbf H_{ls}^{H}$.
Since the APs and BD-RISs are stationary, \(\mathbf H_{ls}\) is modeled as a deterministic line of sight (LoS) channel~\cite{alnahhas2024}, while spatially correlated Rician fading is assumed for the remaining channels~\cite{cha2026}.
The direct user $k$-AP $l$ uplink channel is
{\small
\[
\begin{aligned}
\mathbf f_{lk}
&=
\bar{\mathbf f}_{lk}
+
\mathbf R_{f_{lk}}^{1/2}\tilde{\mathbf f}_{lk},
\qquad
\tilde{\mathbf f}_{lk}\sim\mathcal{CN}(\mathbf0,\mathbf I_M),
\\
\bar{\mathbf f}_{lk}
&=
\sqrt{\frac{M\kappa_{lk}^{f}\beta_{lk}}
{\kappa_{lk}^{f}+1}}\,
\mathbf a_{\mathrm{AP}}(\phi_{lk}^{f}),
\qquad
\mathbf R_{f_{lk}}
=
\frac{\beta_{lk}}
{\kappa_{lk}^{f}+1}\boldsymbol{\Delta}_{lk},
\end{aligned}
\]
}
where $\beta_{lk}$ and $\kappa_{lk}^{f}$ denote the large-scale fading coefficient and Rician factor, respectively, $\phi_{lk}^{f}$ is the angle of arrival (AoA) of the LoS component at AP $l$, and $\mathbf a_{\mathrm{AP}}(\phi_{lk}^{f})$ is the corresponding unit-norm AP array response vector, i.e., $\|\mathbf a_{\mathrm{AP}}(\phi_{lk}^{f})\|^2=1$; moreover, $\operatorname{tr}(\boldsymbol{\Delta}_{lk})=M$.
Similarly, the user $k$-RIS $s$ uplink channel is
{\small
\[
\begin{aligned}
\mathbf z_{sk}
&=
\bar{\mathbf z}_{sk}
+
\mathbf R_{z_{sk}}^{1/2}\tilde{\mathbf z}_{sk},
\qquad
\tilde{\mathbf z}_{sk}\sim\mathcal{CN}(\mathbf0,\mathbf I_N),
\\
\bar{\mathbf z}_{sk}
&=
\sqrt{\frac{N\kappa_{sk}^{z}\tilde{\beta}_{sk}}
{\kappa_{sk}^{z}+1}}\,
\mathbf a_{\mathrm{RIS},s}
(\phi_{sk}^{z},\varphi_{sk}^{z}),
\qquad
\mathbf R_{z_{sk}}
=
\frac{\tilde{\beta}_{sk}}
{\kappa_{sk}^{z}+1}\boldsymbol{\Gamma},
\end{aligned}
\]
}
where the large-scale fading coefficient and Rician factor, are $\tilde{\beta}_{sk}$ and $\kappa_{sk}^{z}$ respectively, $\phi_{sk}^{z}$ and $\varphi_{sk}^{z}$ are the azimuth and elevation AoA at BD-RIS $s$, and $\mathbf a_{\mathrm{RIS},s}(\phi_{sk}^{z},\varphi_{sk}^{z})$ is the corresponding unit-norm array response vector, i.e., $\|\mathbf a_{\mathrm{RIS},s}(\phi_{sk}^{z},\varphi_{sk}^{z})\|^2=1$.
Following~\cite{emilsinc}, we use the normalized isotropic half-space
correlation model
\(
[\boldsymbol{\Gamma}]_{n,m}
=
\operatorname{sinc}(2\|\mathbf q_n-\mathbf q_m\|/\lambda)
\),
where $\mathbf q_n$ is the position of element $n$,
$\lambda$ is the operating wavelength, and
$\operatorname{tr}(\boldsymbol{\Gamma})=N$.
Here, $\boldsymbol{\Gamma}$ models spatial correlation among BD-RIS elements.
The resulting uplink aggregate channel is
\(
\mathbf u_{lk}
=
\mathbf f_{lk}
+
\sum_{s=1}^{S}
\mathbf H_{ls}^{H}\boldsymbol{\Theta}_s\mathbf z_{sk}.
\label{eq:u-lk}
\)
We assume $\{\tilde{\mathbf f}_{lk}\}$ are mutually independent across
$l,k$, $\{\tilde{\mathbf z}_{sk}\}$ are mutually independent across
$s,k$, and the two families are mutually independent. Hence,
$\mathbf u_{lk}\sim
\mathcal{CN}(\bar{\mathbf u}_{lk},\mathbf R_{u_{lk}})$, where
\(
\bar{\mathbf u}_{lk}
=
\bar{\mathbf f}_{lk}
+
\sum_{s=1}^{S}
\mathbf H_{ls}^{H}\boldsymbol{\Theta}_s
\bar{\mathbf z}_{sk},
\)
and
\(
\mathbf R_{u_{lk}}
=
\mathbf R_{f_{lk}}
+
\sum_{s=1}^{S}
\mathbf H_{ls}^{H}\boldsymbol{\Theta}_s
\mathbf R_{z_{sk}}
\boldsymbol{\Theta}_s^{H}\mathbf H_{ls}.
\)
%
%
%
\subsection{Uplink Training Transmission and Aggregate Channel Estimation}
\label{subsec:uplink-training-channel-estimation}
Let $\{\boldsymbol{\psi}_{\pi}\}_{\pi=1}^{\tau_p}$ denote $\tau_p$ orthogonal pilots satisfying
$\|\boldsymbol{\psi}_{\pi}\|^2=\tau_p$ 
and
$\boldsymbol{\psi}_{\pi}^{H}\boldsymbol{\psi}_{\pi'}=0$ 
for
$\pi\neq\pi'$. User $k$ is assigned pilot $\pi_k$, and
$\mathscr P_{\pi}\triangleq\{i:\pi_i=\pi\}$ denotes the corresponding
pilot-sharing set of users. The received signal at AP $l$ during uplink training transmission is
\(
\mathbf Y_l^{\rm p}
=
\sum_{i=1}^{K}
\sqrt{p_i}\,\mathbf u_{li}\boldsymbol{\psi}_{\pi_i}^{H}
+
\mathbf N_l^{\rm p}.
\)
Correlating $\mathbf Y_l^{\rm p}$ with corresponding pilot results in
\(
\mathbf y_{l\pi}^{\rm p}
\triangleq
\frac{1}{\sqrt{\tau_p}}
\mathbf Y_l^{\rm p}\boldsymbol{\psi}_{\pi}
=
\sum_{i\in\mathscr P_{\pi}}
\sqrt{p_i\tau_p}\,\mathbf u_{li}
+
\mathbf n_{l\pi}^{\rm p},
\)
where the entries of $\mathbf N_l^{\rm p}$ are i.i.d.
$\mathcal{CN}(0,\sigma_{\rm ul}^{2})$ and
$\mathbf n_{l\pi}^{\rm p}\sim
\mathcal{CN}(\mathbf0,\sigma_{\rm ul}^{2}\mathbf I_M)$.
Therefore,
{\small
\begin{equation}
\boldsymbol{\Omega}_{l\pi}
\triangleq
\operatorname{Cov}\!\left(\mathbf y_{l\pi}^{\rm p}\right)
=
\tau_p
\sum_{i\in\mathscr P_{\pi}}
p_i\mathbf R_{u_{li}}
+
\sigma_{\rm ul}^{2}\mathbf I_M .
\label{eq: omega}
\end{equation}
}
For $k\in\mathscr K_l$, joint Gaussianity of
$\mathbf u_{lk}$ and $\mathbf y_{l\pi_k}^{\rm p}$ yields the MMSE
 estimator of the uplink aggregate channel
{\small
\begin{equation}
\begin{aligned}
\hat{\mathbf u}_{lk}
&=
\bar{\mathbf u}_{lk}
+
\mathbf W_{lk}
\left(
\mathbf y_{l\pi_k}^{\rm p}
-
\mathbb E\!\left\{\mathbf y_{l\pi_k}^{\rm p}\right\}
\right),
\\
\mathbf W_{lk}
&=
\sqrt{p_k\tau_p}\,
\mathbf R_{u_{lk}}
\boldsymbol{\Omega}_{l\pi_k}^{-1},
\\
\mathbf R_{\hat u_{lk}}
&\triangleq
\operatorname{Cov}\!\left(\hat{\mathbf u}_{lk}\right)
=
p_k\tau_p
\mathbf R_{u_{lk}}
\boldsymbol{\Omega}_{l\pi_k}^{-1}
\mathbf R_{u_{lk}} .
\end{aligned}
\label{eq:mmse-estimate-u-lk}
\end{equation}
}
The estimation error covariance is
\(
\operatorname{Cov}\!\left(
\mathbf u_{lk}-\hat{\mathbf u}_{lk}
\right)
=
\mathbf R_{u_{lk}}-\mathbf R_{\hat u_{lk}}.
\)
Hence, the NMSE is
{\small
\begin{equation}
\zeta_{lk}
\triangleq
\frac{
\operatorname{tr}\!\left(
\mathbf R_{u_{lk}}-\mathbf R_{\hat u_{lk}}
\right)
}{
D_{lk}
},
\ 
D_{lk}
\triangleq
\|\bar{\mathbf u}_{lk}\|^2
+
\operatorname{tr}(\mathbf R_{u_{lk}}),
\ 
k\in\mathscr K_l .
\label{eq:nmse}
\end{equation}
}
%
%
%
\subsection{Statistical Characterization of the Aggregate Channel}
The uplink SINR analysis requires second-order statistics of the uplink aggregate channels, received pilot signals, and channel estimates. The following lemmas do this.
\begin{lemma}
\label{lem:aggregate-channel-second-moments}
Define
\(
\mathbf C_{lk,mi}
\triangleq
\operatorname{Cov}(\mathbf u_{lk},\mathbf u_{mi})
\).
Then,
{\small
\[
\mathbf C_{lk,mi}
=
\begin{cases}
\mathbf R_{u_{lk}},
& m=l,\ i=k,
\\[1mm]
\displaystyle
\sum_{s=1}^{S}
\mathbf H_{ls}^{H}\boldsymbol{\Theta}_s
\mathbf R_{z_{sk}}
\boldsymbol{\Theta}_s^{H}\mathbf H_{ms},
& m\neq l,\ i=k,
\\[2mm]
\mathbf 0,
& i\neq k.
\end{cases}
\]
}
\end{lemma}
\begin{lemma}
\label{lem:pilot-and-estimate-second-moments}
Define the cross covariance 
\(
\mathbf Q_{lk,mi}
\triangleq
\operatorname{Cov}
\!\left(
\hat{\mathbf u}_{lk},
\mathbf u_{mi}
\right).
\)
{\small
\[
\mathbf Q_{lk,mi}
=
\begin{cases}
\tau_p\sqrt{p_kp_i}\,
\mathbf R_{u_{lk}}
\boldsymbol{\Omega}_{l\pi_k}^{-1}
\mathbf C_{li,mi},
& i\in\mathscr P_{\pi_k},
\\[1mm]
\mathbf 0,
& i\notin\mathscr P_{\pi_k}.
\end{cases}
\]
}
The cross covariance between two received pilot signals is
{\small
\[
\begin{aligned}
\boldsymbol{\Omega}^{\rm p}_{l\pi,m\pi'}
&\triangleq
\operatorname{Cov}
\!\left(
\mathbf y_{l\pi}^{\rm p},
\mathbf y_{m\pi'}^{\rm p}
\right)
\\
&=
\delta_{\pi\pi'}
\left(
\tau_p
\sum_{j\in\mathscr P_{\pi}}
p_j\mathbf C_{lj,mj}
+
\delta_{lm}\sigma_{\rm ul}^{2}\mathbf I_M
\right),
\end{aligned}
\]
}
where $\delta_{ab}$ denotes the Kronecker delta. Hence, the
cross covariance between two channel estimates is
{\small
\[
\mathbf T_{lk,mi}
\triangleq
\operatorname{Cov}
\!\left(
\hat{\mathbf u}_{lk},
\hat{\mathbf u}_{mi}
\right)
=
\mathbf W_{lk}
\boldsymbol{\Omega}^{\rm p}_{l\pi_k,m\pi_i}
\mathbf W_{mi}^{H}.
\]
}
Note that,
\(
\boldsymbol{\Omega}^{\rm p}_{l\pi,l\pi}
=
\boldsymbol{\Omega}_{l\pi}
\)
and
\(
\mathbf Q_{lk,lk}
=
\mathbf T_{lk,lk}
=
\mathbf R_{\hat u_{lk}},
\)
while
\(
\mathbf Q_{lk,mi}
=
\mathbf T_{lk,mi}
=
\mathbf 0
\)
for $\pi_i\neq\pi_k$.
\end{lemma}
%
%
%
\section{Uplink Data Transmission}
\label{sec:uplink-data}
After uplink training, user $i$ transmits the normalized symbol $s_i$
with power $\rho_i$, where $\mathbb E\{|s_i|^2\}=1$,
$\mathbb E\{s_i s_j^*\}=0$ for $i\neq j$, and
$0\leq\rho_i\leq\rho_{\max}$. The received signal at AP $l$ is
{\small
\begin{equation}
    \mathbf y_l
    =
    \sum_{i=1}^{K}\sqrt{\rho_i}\,\mathbf u_{li}s_i
    +
    \mathbf n_l ,
    \qquad
    \mathbf n_l \sim
    \mathcal{CN}(\mathbf 0,\sigma_{\rm ul}^{2}\mathbf I_M).
\end{equation}
}
The APs forward their received uplink signals and the corresponding channel estimates to the CPU over the error-free fronthaul links. For user \(k\), the CPU then performs MR combining using the signals and channel estimates from the serving APs in \(\mathscr L_k\), as
\(
    \hat{s}_k
    =
    \sum_{l\in\mathscr L_k}
    \hat{\mathbf u}_{lk}^{H}\mathbf y_l .
\)
For tractability, we define
\(
a_{ki}\triangleq
\sum_{l\in\mathscr L_k}
\hat{\mathbf u}_{lk}^{H}\mathbf u_{li}.
\)
Then,
{\small
\begin{equation}
\hat{s}_k
=
\mathrm{DS}_k s_k
+
\mathrm{BU}_k s_k
+
\sum\nolimits_{\substack{i=1\\i\neq k}}^{K} \mathrm{UI}_{ki} s_i
+
\mathrm{N}_k,
\label{eq:ul-uatf-decomposition}
\end{equation}
}
where
{\small
\begin{align*}
\mathrm{DS}_k
&\triangleq
\sqrt{\rho_k}\,\mathbb E\{a_{kk}\},
&
\mathrm{BU}_k
&\triangleq
\sqrt{\rho_k}\!
\left(a_{kk}-\mathbb E\{a_{kk}\}\right),
\\
\mathrm{UI}_{ki}
&\triangleq
\sqrt{\rho_i}\,a_{ki},
&
\mathrm{N}_k
&\triangleq
\sum_{l\in\mathscr L_k}
\hat{\mathbf u}_{lk}^{H}\mathbf n_l.
\end{align*}
}
Treating all terms in \eqref{eq:ul-uatf-decomposition} except $\mathrm{DS}_ks_k$ as effective noise yields
{\small
\begin{equation}
    \gamma_k 
    =
    \frac{
    |\mathrm{DS}_k|^2
    }{
    \mathbb E\{|\mathrm{BU}_k|^2\}
    +
    \sum_{\substack{i=1\\i\neq k}}^{K}
    \mathbb E\{|\mathrm{UI}_{ki}|^2\}
    +
    \mathbb E\{|\mathrm{N}_k|^2\}
    }.
    \label{eq:ul-sinr-general}
\end{equation}
}
To reveal the effect of pilot reuse, define the aggregate pilot-sharing
and non-pilot-sharing interference powers as
{\small
\[
\mathcal I_{k}^{\rm PS}
\triangleq
\sum_{i\in\mathscr P_{\pi_k}\setminus\{k\}}
\mathbb E\!\left\{|\mathrm{UI}_{ki} |^2\right\},
\qquad
\mathcal I_{k}^{\rm NPS}
\triangleq
\sum_{i\notin\mathscr P_{\pi_k}}
\mathbb E\!\left\{|\mathrm{UI}_{ki} |^2\right\}.
\]
}
Hence,
{\small
\(
\sum\nolimits_{\substack{i=1\\ i\neq k}}^{K}
\mathbb E\!\left\{|\mathrm{UI}_{ki} |^2\right\}
=
\mathcal I_{k}^{\rm PS}
+
\mathcal I_{k}^{\rm NPS}.
\)
}
The corresponding achievable uplink ergodic SE is lower bounded by
\(
    \mathrm{SE}_k
    =
    \tau_u/\tau_c
    \log_2\left(1+\gamma_k \right).
\)
%
\begin{theorem}
\label{thm:closed-form-uplink-sinr}
Under MR combining, define
\(
\mu_{lk}
\triangleq
\|\bar{\mathbf u}_{lk}\|^2
+
\operatorname{tr}(\mathbf R_{\hat u_{lk}})
\)
and
{\small
\begin{align*}
\mathcal V_{ki}
\triangleq
\sum_{l\in\mathscr L_k}
\sum_{m\in\mathscr L_k}
\Big(
&
\bar{\mathbf u}_{lk}^{H}
\mathbf C_{li,mi}
\bar{\mathbf u}_{mk}
+
\bar{\mathbf u}_{mi}^{H}
\mathbf T_{mk,lk}
\bar{\mathbf u}_{li}
\\
&+
\operatorname{tr}
(\mathbf T_{mk,lk}\mathbf C_{li,mi})
\Big).
\end{align*}
}
Then, the terms in \eqref{eq:ul-sinr-general} are given by
{\small
\begin{align}
|\mathrm{DS}_k |^2
&=
\rho_k
\left|
\sum_{l\in\mathscr L_k}\mu_{lk}
\right|^2,
\qquad
\mathbb E\{|\mathrm{BU}_k |^2\}
=
\rho_k\mathcal V_{kk},
\nonumber
\\
\mathcal I_k^{\rm PS}
&=
\sum_{i\in\mathscr P_{\pi_k}\setminus\{k\}}
\!\!\!\!\!\!\!\rho_i
\left[
\left|
\sum_{l\in\mathscr L_k}
\left(
\bar{\mathbf u}_{lk}^{H}\bar{\mathbf u}_{li}
+
\operatorname{tr}(\mathbf Q_{lk,li}^{H})
\right)
\right|^2
+
\mathcal V_{ki}
\right],
\nonumber
\\
\mathcal I_k^{\rm NPS}
&=
\sum_{i\notin\mathscr P_{\pi_k}}
\rho_i
\left[
\left|
\sum_{l\in\mathscr L_k}
\bar{\mathbf u}_{lk}^{H}\bar{\mathbf u}_{li}
\right|^2
+
\mathcal V_{ki}
\right],
\nonumber
\\
\mathbb E\{|\mathrm N_k |^2\}
&=
\sigma_{\rm ul}^{2}
\sum_{l\in\mathscr L_k}\mu_{lk}.
\nonumber
\end{align}
}
\end{theorem}
The proofs of Lemmas~1--2 and Theorem~1 are omitted due to space
limitations.
%
%
%
\section{BD-RIS Scattering Matrix Optimization}
\label{sec:bd-ris-optimization}
\subsection{Problem Formulation and Riemannian Optimization}
\label{subsec:problem-formulation}
Let
$\boldsymbol{\Theta}\triangleq
\{\boldsymbol{\Theta}_{s,g}\}_{s=1,g=1}^{S,G}$
denote the collection of BD-RIS scattering blocks. The resulting
optimization problem is
{\small
\begin{align}
    \underset{\boldsymbol{\Theta}}{\operatorname{minimize}}
    \quad &
    f(\boldsymbol{\Theta})
    \triangleq
    \sum_{k=1}^{K}
    \sum_{l\in\mathscr L_k}
    \zeta_{lk}
    \label{prob:bd-ris-nmse}
    \\
    \operatorname{subject\ to}\quad
    &
    \boldsymbol{\Theta}_{s,g}
    =
    \boldsymbol{\Theta}_{s,g}^{T},
    \qquad \forall s,g,
    \nonumber\\
    &
    \boldsymbol{\Theta}_{s,g}^{H}
    \boldsymbol{\Theta}_{s,g}
    =
    \mathbf I_{\bar N},
    \qquad \forall s,g.
    \nonumber
\end{align}
}
%
%
%
We solve \eqref{prob:bd-ris-nmse} using Riemannian steepest descent with Armijo backtracking over the feasible product manifold. The main difficulty is that each scattering block must remain simultaneously symmetric and unitary, so a standard Euclidean update does not generally preserve feasibility. The symmetric unitary manifold geometry of an individual scattering block was characterized in \cite{santamaria2026manifold}, while the product manifold formulation across the groups of a group-connected BD-RIS was developed in \cite{fidanovski2026}. We adopt this geometric framework and extend the product construction to the $S$ BD-RISs considered here, resulting in $SG$ jointly optimized scattering blocks coupled through the aggregate channel statistics.
For one BD-RIS group, define
\(
    \mathscr M
    \triangleq
    \left\{
    \mathbf X\in\mathbb C^{\bar N\times\bar N}
    \mid
    \mathbf X=\mathbf X^T,\;
    \mathbf X^H\mathbf X=\mathbf I_{\bar N}
    \right\}.
\)
Thus, $\boldsymbol{\Theta}_{s,g}\in\mathscr M$ for all $s,g$, and the
joint feasible set is
\(
    \mathscr M_{\mathrm{prod}}
    \triangleq
    \prod_{s=1}^{S}\prod_{g=1}^{G}\mathscr M.
\)
Accordingly, \eqref{prob:bd-ris-nmse} can be equivalently written as
\(
    \min_{\boldsymbol{\Theta}\in\mathscr M_{\mathrm{prod}}}
    f(\boldsymbol{\Theta}).
\)
We equip each manifold factor with the metric induced by the real
Euclidean inner product
\(
    \langle\mathbf X,\mathbf Y\rangle
    \triangleq
    \Re\!\left\{
    \operatorname{tr}(\mathbf X^H\mathbf Y)
    \right\},
\)
and $\mathscr M_{\mathrm{prod}}$ with the corresponding product metric. 
The geometric operations are applied factorwise, whereas the Euclidean
gradient derived below captures the coupling induced by the proposed multi-BD-RIS statistical sum-NMSE objective.
Let
$\mathbf G_{s,g}\triangleq
\nabla_{\boldsymbol{\Theta}_{s,g}}f$
denote the Euclidean gradient under the adopted metric, defined through
\(
df=
\Re\!\left\{
\operatorname{tr}
\left(
\mathbf G_{s,g}^{H}d\boldsymbol{\Theta}_{s,g}
\right)
\right\}.
\)
To obtain its explicit form, we partition the BD-RIS-related channel
quantities according to the $G$ groups.
Consistently with the block-diagonal structure of $\boldsymbol{\Theta}_s$, write
\(
    \mathbf H_{ls}
    =
    \begin{bmatrix}
        \mathbf H_{ls,1}^{T} &
        \cdots &
        \mathbf H_{ls,G}^{T}
    \end{bmatrix}^{T},
    \ 
    \mathbf H_{ls,g}\in\mathbb C^{\bar N\times M},
\)
and
\(
    \mathbf z_{sk}
    =
    \begin{bmatrix}
        \mathbf z_{sk,1}^{T} &
        \cdots &
        \mathbf z_{sk,G}^{T}
    \end{bmatrix}^{T},
    \
    \bar{\mathbf z}_{sk}
    =
    \begin{bmatrix}
        \bar{\mathbf z}_{sk,1}^{T} &
        \cdots &
        \bar{\mathbf z}_{sk,G}^{T}
    \end{bmatrix}^{T},
\)
where
$\mathbf z_{sk,g},\bar{\mathbf z}_{sk,g}\in\mathbb C^{\bar N}$.
The covariance matrix is conformably partitioned as
\(
\mathbf R_{z_{sk}}
=
[\mathbf R_{z_{sk}}^{(g,h)}]_{g,h=1}^{G},
\quad
\mathbf R_{z_{sk}}^{(g,h)}
\in\mathbb C^{\bar N\times\bar N}.
\)
In general, the off-diagonal blocks
$\mathbf R_{z_{sk}}^{(g,h)}$, $g\neq h$, are nonzero due to spatial
correlation across the BD-RIS aperture. Since $\mathbf R_{z_{sk}}$
is proportional to $\boldsymbol{\Gamma}$, it is
partitioned into the same $G\times G$ block structure.
where
\(
    \mathbf R_{z_{sk}}^{(g,h)}
    \triangleq
    \mathbb E
    \left\{
    \tilde{\mathbf z}_{sk,g}
    \tilde{\mathbf z}_{sk,h}^{H}
    \right\}
    \in\mathbb C^{\bar N\times\bar N}.
\)
Since $\mathbf R_{z_{sk}}$ is proportional to
$\boldsymbol{\Gamma}$ under the adopted channel model,
$\boldsymbol{\Gamma}$ is partitioned into the same $G\times G$
block structure.
For a perturbation of the $(s,g)$-th scattering block, the variation
of the aggregate channel mean is
\(
    d\bar{\mathbf u}_{lk}
    =
    \mathbf H_{ls,g}^{H}
    d\boldsymbol{\Theta}_{s,g}
    \bar{\mathbf z}_{sk,g}.
\)
For compactness, define the auxiliary quantities
{\small
\begin{equation}
\begin{aligned}
\mathbf B_{lsi,g}
&\triangleq
\sum_{h=1}^{G}
\mathbf R_{z_{si}}^{(g,h)}
\boldsymbol{\Theta}_{s,h}^{H}
\mathbf H_{ls,h},
\\
\boldsymbol{\Xi}_{lk}
&\triangleq
(1-\zeta_{lk})\mathbf I_M
-
p_k\tau_p
\left(
\boldsymbol{\Omega}_{l\pi_k}^{-1}\mathbf R_{u_{lk}}
+
\mathbf R_{u_{lk}}\boldsymbol{\Omega}_{l\pi_k}^{-1}
\right),
\\
\boldsymbol{\Psi}_{lk}
&\triangleq
p_k\tau_p\,
\boldsymbol{\Omega}_{l\pi_k}^{-1}
\mathbf R_{u_{lk}}^{2}
\boldsymbol{\Omega}_{l\pi_k}^{-1}.
\end{aligned}
\label{eq:gradient-auxiliary}
\end{equation}
}
Using
$\mathbf R_{z_{si}}^{(h,g)}
=(\mathbf R_{z_{si}}^{(g,h)})^{H}$,
the corresponding covariance variation is
\(
d\mathbf R_{u_{li}}
=
\mathbf H_{ls,g}^{H}
d\boldsymbol{\Theta}_{s,g}
\mathbf B_{lsi,g}
+
\mathbf B_{lsi,g}^{H}
d\boldsymbol{\Theta}_{s,g}^{H}
\mathbf H_{ls,g}.
\)
Using the first-order variations of \(\bar{\mathbf u}_{lk}\), \(\mathbf R_{u_{lk}}\), and \(\boldsymbol{\Omega}_{l\pi_k}^{-1}\), together with the cyclic trace identity, the Euclidean gradient with respect to the \((s,g)\)-th scattering block is obtained as
{\small
\setlength{\jot}{1pt}
\begin{equation}
\begin{aligned}
\mathbf G_{s,g}
&=
2\sum_{k=1}^{K}\sum_{l\in\mathscr L_k}
\frac{1}{D_{lk}}\,
\mathbf H_{ls,g}
\Bigg[
\boldsymbol{\Xi}_{lk}\mathbf B_{lsk,g}^{H}
\\[-1mm]
&\qquad
+
\boldsymbol{\Psi}_{lk}
\sum_{i\in\mathscr P_{\pi_k}}
p_i\tau_p\,\mathbf B_{lsi,g}^{H}
-
\zeta_{lk}
\bar{\mathbf u}_{lk}
\bar{\mathbf z}_{sk,g}^{H}
\Bigg].
\end{aligned}
\label{eq:euclidean-gradient-G-sg}
\end{equation}
}
At a feasible $\boldsymbol{\Theta}_{s,g}$, differentiating the symmetry and unitary constraints in \eqref{prob:bd-ris-nmse} gives the tangent space
{\small
\begin{equation}
    \!\!\!\!\mathscr T_{\boldsymbol{\Theta}_{s,g}}
    =
    \left\{
    \mathbf A\in\mathbb C^{\bar N\times\bar N}
    \,\middle|\,
    \mathbf A=\mathbf A^T,\;
    \boldsymbol{\Theta}_{s,g}^{H}\mathbf A
    +
    \mathbf A^{H}\boldsymbol{\Theta}_{s,g}
    =
    \mathbf 0
    \right\}.
    \label{eq:tangent-space}
\end{equation}
}
For each feasible scattering block, let
\(
    \boldsymbol{\Theta}_{s,g}
    =
    \mathbf V_{s,g}\mathbf V_{s,g}^{T}
\)
denote a Takagi factorization, where $\mathbf V_{s,g}$ is unitary.
Following \cite{santamaria2026manifold}, under the adopted metric the Riemannian gradient is obtained by orthogonally projecting $\mathbf G_{s,g}$ onto
$\mathscr T_{\boldsymbol{\Theta}_{s,g}}$ as
{\small
\begin{equation}
\begin{aligned}
\operatorname{grad}_{\boldsymbol{\Theta}_{s,g}} f
&\triangleq
\mathcal P_{\mathscr T_{\boldsymbol{\Theta}_{s,g}}}
\!\left(\mathbf G_{s,g}\right)
\\
&=
j\mathbf V_{s,g}
\Im\!\left\{
\mathbf V_{s,g}^{H}
\frac{\mathbf G_{s,g}+\mathbf G_{s,g}^{T}}{2}
\mathbf V_{s,g}^{*}
\right\}
\mathbf V_{s,g}^{T}.
\end{aligned}
\label{eq:tangent-projection}
\end{equation}
}
At iteration $t$, the blockwise Riemannian steepest-descent direction is
\(
    \boldsymbol{\xi}_{s,g}^{(t)}
    \triangleq
    -
    \operatorname{grad}_{\boldsymbol{\Theta}_{s,g}^{(t)}} f,
    \
    \boldsymbol{\xi}_{s,g}^{(t)}
    \in
    \mathscr T_{\boldsymbol{\Theta}_{s,g}^{(t)}}.
\)
For a trial step size $\nu>0$, the symmetric blockwise update admits a
Takagi factorization. Following
\cite{santamaria2026manifold,fidanovski2026}, we use the corresponding
symmetric unitary retraction
{\small
\begin{equation}
\begin{aligned}
&\mathbf Y_{s,g}^{(t)}(\nu)
=
\boldsymbol{\Theta}_{s,g}^{(t)}
+\nu\boldsymbol{\xi}_{s,g}^{(t)}
=
\mathbf U_{s,g}^{(t)}
\boldsymbol{\Sigma}_{s,g}^{(t)}
\bigl(\mathbf U_{s,g}^{(t)}\bigr)^T,
\\
&\mathcal R_{\boldsymbol{\Theta}_{s,g}^{(t)}}
\!\left(\nu\boldsymbol{\xi}_{s,g}^{(t)}\right)
=
\mathbf U_{s,g}^{(t)}
\bigl(\mathbf U_{s,g}^{(t)}\bigr)^T.
\end{aligned}
\label{eq:takagi-retraction}
\end{equation}
}
where the retraction replaces the Takagi singular values by unity,
thereby preserving the symmetric unitary constraints. The retraction is
applied independently to all $SG$ factors of
$\mathscr M_{\mathrm{prod}}$, with
$\boldsymbol{\xi}^{(t)}
\triangleq
\{\boldsymbol{\xi}_{s,g}^{(t)}\}_{s=1,g=1}^{S,G}$.
The step size is selected by Armijo backtracking as
\(
\nu^{(t)}
=
\nu_0\chi^{r_t},
\ 0<\chi<1,
\ 0<c<1.
\)
Define the squared norm of the product-manifold Riemannian gradient as
\(
    \eta^{(t)}
    \triangleq
    \sum_{s=1}^{S}
    \sum_{g=1}^{G}
    \left\|
    \operatorname{grad}_{\boldsymbol{\Theta}_{s,g}^{(t)}}f
    \right\|_F^2.
\)
{\small
\begin{equation}
f\!\left(
\mathcal R_{\boldsymbol{\Theta}^{(t)}}
\!\left(
\nu^{(t)}\boldsymbol{\xi}^{(t)}
\right)
\right)
\le
f\!\left(\boldsymbol{\Theta}^{(t)}\right)
-c\nu^{(t)}
\eta^{(t)}
\label{eq:armijo}
\end{equation}
}
Here, $r_t$ is the smallest nonnegative integer for which
\eqref{eq:armijo} holds. The accepted retracted point defines
$\boldsymbol{\Theta}^{(t+1)}$. The complete Riemannian optimization procedure is summarized in Algorithm~\ref{alg:bd-ris-rgd}.
\begin{algorithm}[h]
\footnotesize
\caption{Long-Term Riemannian BD-RIS Optimization}
\label{alg:bd-ris-rgd}
\begin{algorithmic}[1]

\State \textbf{Input:}
statistics, pilots, feasible
$\boldsymbol{\Theta}^{(0)}\in\mathscr M_{\mathrm{prod}}$,
$\nu_0,\chi,c,\epsilon,T_{\max}$

\For{$t=0,\ldots,T_{\max}-1$}

    \State Evaluate $f(\boldsymbol{\Theta}^{(t)})$ and the quantities
    required by \eqref{eq:euclidean-gradient-G-sg}

    \State Compute
    $\operatorname{grad}_{\boldsymbol{\Theta}_{s,g}^{(t)}}f$
    and
    $\boldsymbol{\xi}_{s,g}^{(t)}
    =-\operatorname{grad}_{\boldsymbol{\Theta}_{s,g}^{(t)}}f$,
    $\forall s,g$

    \State Compute $\eta^{(t)}$

    \If{$\eta^{(t)}\leq\epsilon$}
        \State \textbf{break}
    \EndIf

    \State Select $\nu^{(t)}=\nu_0\chi^{r_t}$ using
    \eqref{eq:armijo}

    \State Update $\boldsymbol{\Theta}^{(t+1)}$ using
    \eqref{eq:takagi-retraction}

\EndFor

\State \textbf{Output:} $\boldsymbol{\Theta}^{(t)}$

\end{algorithmic}
\end{algorithm}
\subsection{Convergence Analysis}
\label{subsec:convergence-analysis}
We first note that the Takagi retraction in
\eqref{eq:takagi-retraction} preserves the symmetric unitary constraints.
Hence, starting from a feasible
$\boldsymbol{\Theta}^{(0)}$, all subsequent iterates remain in
$\mathscr M_{\mathrm{prod}}$. Moreover, each feasible block satisfies
$\|\boldsymbol{\Theta}_{s,g}\|_F^2=\bar N$, so the finite product
$\mathscr M_{\mathrm{prod}}$ is compact. The objective is well defined
and continuously differentiable on this set since
$\boldsymbol{\Omega}_{l\pi_k}
\succeq \sigma_{\rm ul}^2\mathbf I_M\succ\mathbf 0$
for every $l,k$, and $D_{lk}>0$ for all served AP-user links, as holds whenever
the corresponding aggregate channel has nonzero average power.
Since
$\boldsymbol{\xi}_{s,g}^{(t)}
=
-\operatorname{grad}_{\boldsymbol{\Theta}_{s,g}^{(t)}}f$,
the directional derivative under the product metric is
$-\eta^{(t)}<0$ whenever $\eta^{(t)}>0$. Hence, the Armijo condition
\eqref{eq:armijo} gives
{\small
\(
    f\!\left(\boldsymbol{\Theta}^{(t+1)}\right)
    \leq
    f\!\left(\boldsymbol{\Theta}^{(t)}\right)
    -
    c\nu^{(t)}\eta^{(t)}
    \leq
    f\!\left(\boldsymbol{\Theta}^{(t)}\right).
    \label{eq:objective-descent}
\)
}
Since each NMSE is nonnegative,
$f(\boldsymbol{\Theta})\geq0$; therefore,
$\{f(\boldsymbol{\Theta}^{(t)})\}$ is monotonically nonincreasing,
bounded below, and converges to a finite value.
By the global convergence result for Riemannian line-search methods
\cite
{absil2008optimization}, every accumulation point of
the generated sequence is a critical point of $f$. Since
$\mathscr M_{\mathrm{prod}}$ is compact and the Riemannian gradient is
continuous, this further implies $\eta^{(t)} \longrightarrow 0$. Thus,
{\small
\(
\left\|
\operatorname{grad}_{\boldsymbol{\Theta}_{s,g}^{(t)}}f
\right\|_F
\longrightarrow 0,
\qquad \forall\,s,g.
\label{eq:gradient-convergence}
\)
}
Therefore, the objective values converge monotonically and the
product-manifold Riemannian gradient vanishes asymptotically. Since
\eqref{prob:bd-ris-nmse} is nonconvex, these results establish
first-order stationarity rather than global optimality.
\subsection{Computational Complexity}
\label{subsec:complexity}
We characterize the dominant per-iteration complexity of Algorithm~\ref{alg:bd-ris-rgd} using worst-case asymptotic upper bounds. The analysis follows line~3 for objective evaluation, line~4 for Riemannian gradient computation, and lines~9--10 for the Armijo search and retraction.
For line~3, the main BD-RIS-dependent covariance term for each AP--BD-RIS pair is
$\mathbf H_{ls}^{H}\boldsymbol{\Theta}_{s}
\boldsymbol{\Gamma}\boldsymbol{\Theta}_{s}^{H}\mathbf H_{ls}$,
whose cost is $\mathcal O(G\bar N^{2}M+N^{2}M+NM^{2})$ by exploiting the group-connected structure of $\boldsymbol{\Theta}_{s}$.
Since $\mathbf R_{z_{sk}}=c_{sk}\boldsymbol{\Gamma}$, these matrix-valued products can be reused across users.
Accounting also for the aggregate channel statistics and the quantities in~\eqref{eq: omega}-\eqref{eq:nmse}, the objective evaluation cost is upper bounded by
{\small
\[
\begin{aligned}
C_f=\mathcal O\Bigg(
&LS\!\left(G\bar N^{2}M+N^{2}M+NM^{2}\right)
+SMN\sum_{l=1}^{L}|\mathscr K_l|
\\
&+LKSM^{2}
+\left(L\tau_p+\sum_{l=1}^{L}|\mathscr K_l|\right)M^{3}
\Bigg).
\end{aligned}
\]
}
Here, the $LK$ term is a worst-case bound on pilot-dependent covariance computations, while at most $\tau_p$ distinct $M\times M$ pilot covariance matrices are inverted per AP.
In line~4, the quantities above are reused in~\eqref{eq:euclidean-gradient-G-sg}, giving
\(
C_{\nabla}
=\mathcal O\!\left(
SG\sum_{l=1}^{L}|\mathcal K_l|
\left(\bar N M^{2}+\bar N^{2}M\right)
+SG\bar N^{3}
\right),
\)
where the last term accounts for the factorwise tangent space projection in~\eqref{eq:tangent-projection}.
Finally, each Armijo trial in lines~9--10 requires one objective evaluation and $SG$ Takagi retractions, with retraction cost $\mathcal O(SG\bar N^{3})$. Thus, if $r_t$ is the number of backtracking reductions at iteration $t$,
\(
C^{(t)}
=
C_f+C_{\nabla}
+(r_t+1)\left(C_f+\mathcal O(SG\bar N^{3})\right).
\)
Since $N=G\bar N$, $SG\bar N^{3}=SN^{3}/G^{2}$. Hence, for fixed $N$, increasing $G$ reduces the manifold-related cost, while the dense $\boldsymbol{\Gamma}$ multiplication retains an $\mathcal O(LSN^{2}M)$ contribution.
%
%
%
\section{Numerical Analysis}
\label{sec:numerical-analysis}
\begin{figure}[htbp]
    \vspace{-3mm}
    \centering
    \includegraphics[height=1.5in, keepaspectratio]
    {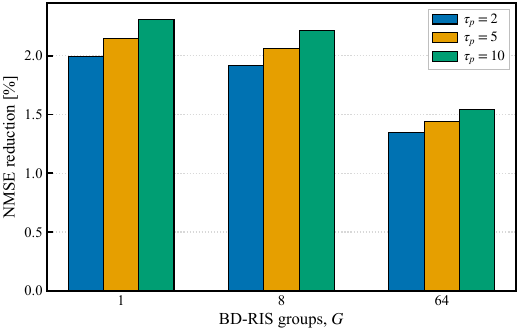}
    \vspace{-3mm}
    \caption{NMSE reduction relative to equal phase versus the number of
             BD-RIS groups $G$ for different pilot lengths $\tau_p$.}
    \label{fig:nmse_reduction_groups}
    \vspace{-3mm}
\end{figure}

\begin{figure}[htbp]
    \vspace{-3mm}
    \centering
    \includegraphics[height=1.5in, keepaspectratio]
    {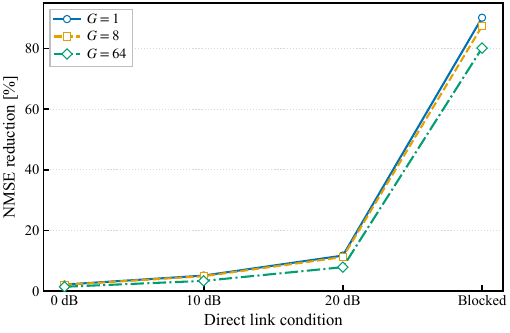}
    \vspace{-3mm}
    \caption{NMSE reduction versus direct-link attenuation for
             $\tau_p=5$.}
    \label{fig:direct-link-sensitivity}
    \vspace{-2.5mm}
\end{figure}

\begin{figure}[htbp]
    \centering
    \includegraphics[height=1.5in, keepaspectratio]
    {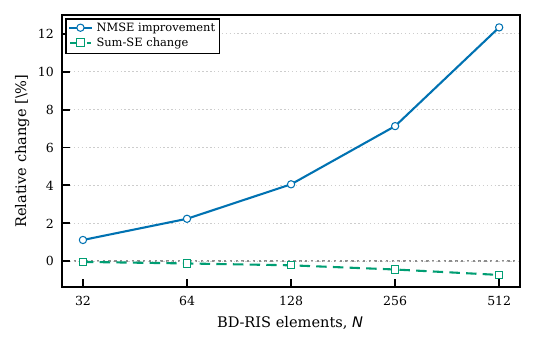}
    \vspace{-5mm}
    \caption{NMSE reduction and sum-SE change versus $N$ for
             $\tau_p=5$ and fixed $\bar N=8$.}
    \label{fig:n-scaling}
    \vspace{-3mm}
\end{figure}

\begin{figure}[htbp]
    \centering
    \includegraphics[height=1.5in, keepaspectratio]{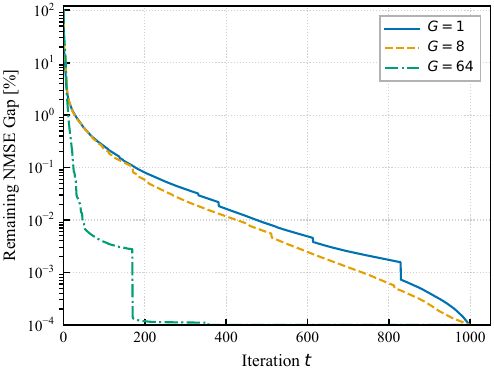}
    \vspace{-3mm}
    \caption{Average remaining objective gap versus iteration for
             $\tau_p=5$.}
    \label{fig:convergence_tau5_semilogy}
    \vspace{-3mm}
\end{figure}

We consider a \(1~\mathrm{km}\times1~\mathrm{km}\) wrap-around area with \(L=20\) APs, \(M=4\) antennas per AP, \(K=10\) single-antenna users, and \(S=2\) BD-RISs located at \((\pm250,0)\) m. APs and users are independently and uniformly deployed with minimum AP--user, AP--BD-RIS, and BD-RIS--user distances of \(10\), \(20\), and \(20\) m, respectively. The AP, BD-RIS, and user heights are \(15\), \(15\), and \(1.65\) m. Unless otherwise stated, each BD-RIS has \(N=64\) elements arranged as an \(8\times8\) half-wavelength rectangular array, while the APs employ half-wavelength ULAs. We use the correlated Rician model of Section~\ref{sec:system-model}, with Rician factors \(3\) dB and \(5\) dB for the AP--user and BD-RIS--user links. The large-scale gain is \(\beta_x(d)[\mathrm{dB}]=\ell_{0,x}-\ell_{1,x}\log_{10}(d/1\,\mathrm{m})+F_x\), where \((\ell_{0,\mathrm{AU}},\ell_{1,\mathrm{AU}})=(-30.5,36.7)\), \((\ell_{0,\mathrm{RIS}},\ell_{1,\mathrm{RIS}})=(-30.62,26)\), and \(F_x\sim\mathcal N(0,4^2)\) dB. The AP-side correlation follows the Gaussian local-scattering model with \(15^\circ\) angular spread, while the BD-RIS correlation follows the isotropic half-space model in Section~\ref{sec:system-model}.
We set \(f_c=2\) GHz, \(B=20\) MHz, and \(\tau_c=200\), and consider \(\tau_p\in\{2,5,10\}\), corresponding to pilot reuse factors \(K/\tau_p\in\{5,2,1\}\). Pilots are assigned through balanced random reuse, with \(p_k=100\) mW and \(\rho_k=\rho_{\max}=200\) mW. The thermal-noise density is \(-174\) dBm/Hz with a \(7\) dB receiver noise figure. Each user is served by its \(L_{\mathrm{srv}}=10\) strongest APs, following~\cite{ezaz2026icc}, where the user-centric performance was observed to saturate around this cluster size. For \(N=64\), we consider \(G\in\{1,8,64\}\), corresponding to \(\bar N\in\{64,8,1\}\) and hence fully connected, group-connected, and single-connected architectures, respectively. The optimizer is initialized from the equal phase configuration with \(\nu_0=10^4\), \(\chi=0.5\), \(c=10^{-4}\), at most \(50\) backtracking trials, and \(T_{\max}=1000\). Unless otherwise stated, results are averaged over $200$ independent large-scale network realizations.
%
To quantify the optimization gain, Fig.~\ref{fig:nmse_reduction_groups} reports the relative NMSE reduction with respect to the equal phase configuration, defined as
\(
\Delta_{\mathrm{NMSE}}
\triangleq
100(\overline{\zeta}_{\mathrm{eq}}-\overline{\zeta}_{\mathrm{opt}})
/
\overline{\zeta}_{\mathrm{eq}}
\),
where \(\overline{\zeta}_{\mathrm{eq}}\) and \(\overline{\zeta}_{\mathrm{opt}}\) denote the average NMSEs under the equal phase and optimized configurations, respectively.
Fig.~\ref{fig:nmse_reduction_groups} shows that the optimized BD-RIS configuration reduces the NMSE for all considered connectivity levels and pilot lengths. The fully connected architecture (\(G=1\)) provides the largest gain, while the group-connected design with \(G=8\) retains nearly all of this benefit with substantially lower inter-element connectivity. For example, at \(\tau_p=5\), the NMSE reductions are \(2.15\%\), \(2.06\%\), and \(1.44\%\) for \(G=1\), \(8\), and \(64\), respectively. The gain also increases moderately with \(\tau_p\) for every \(G\), indicating that the estimation-oriented BD-RIS optimization remains beneficial under pilot reuse, with a slightly larger relative improvement as pilot contamination is reduced.
%
To assess how the direct AP--user link affects the optimization gain, Fig.~\ref{fig:direct-link-sensitivity} shows \(\Delta_{\mathrm{NMSE}}\) for \(\tau_p=5\) as the direct-link power is progressively attenuated.
Fig.~\ref{fig:direct-link-sensitivity} shows that the optimization gain increases sharply as the direct link weakens. For \(G=1\), \(\Delta_{\mathrm{NMSE}}\) rises from \(2.14\%\) under the nominal condition to \(11.70\%\) at \(20\) dB attenuation, with \(G=8\) exhibiting a nearly identical trend. Under complete blockage, the reductions reach \(90.15\%\), \(87.39\%\), and \(80.14\%\) for \(G=1\), \(8\), and \(64\), respectively. These results indicate that the modest nominal gains are largely due to the strong direct AP--user component; as it weakens, the BD-RIS-assisted channel has greater influence on the aggregate channel statistics and hence provides additional degrees of freedom for optimization. The \(G=8\) design again retains most of the fully connected gain with substantially lower inter-element connectivity.
%
We next study the effect of the BD-RIS size while preserving a scalable group-connected architecture. We fix \(\bar N=8\) and vary \(N\in\{32,64,128,256,512\}\), such that \(G=N/\bar N\). Fig.~\ref{fig:n-scaling} reports the results for \(\tau_p=5\), averaged over \(30\) independent network realizations. The relative sum-SE change is defined analogously as
\(
\Delta_{\mathrm{SE}}
\triangleq
100(\mathrm{SE}_{\mathrm{opt}}-\mathrm{SE}_{\mathrm{eq}})
/
\mathrm{SE}_{\mathrm{eq}}
\).
As shown in Fig.~\ref{fig:n-scaling}, \(\Delta_{\mathrm{NMSE}}\) increases from \(1.12\%\) at \(N=32\) to \(12.34\%\) at \(N=512\), with the increasing trend observed in all \(30\) realizations. Thus, enlarging the BD-RIS provides progressively greater control over the  aggregate channel statistics even when the group size is kept fixed.
In contrast, the corresponding sum-SE change remains small and slightly negative, from approximately \(-0.04\%\) to \(-0.73\%\), despite the increasingly large NMSE improvement. This reveals that minimizing the unweighted sum-NMSE does not necessarily improve the uplink SE. Our per-link analysis indicates that most of the NMSE reduction occurs on weaker serving links, whereas MR detection is dominated by the stronger links. Consequently, substantial improvements in aggregate channel-estimation quality may produce little or no SE gain when the optimized links contribute unequally to data detection.
%
Finally, Fig.~\ref{fig:convergence_tau5_semilogy} illustrates the numerical convergence behavior for \(\tau_p=5\). For realization \(n\), define the remaining objective gap as
\(
\mathrm{Gap}_n^{(t)}
\triangleq
100\bigl(f_n^{(t)}-f_n^{(\mathrm{final})}\bigr)/
\bigl(f_n^{(0)}-f_n^{(\mathrm{final})}\bigr)
\),
and the figure reports its average over the network realizations. Thus, the gap starts at \(100\%\) and approaches zero as the iterate approaches its final attained objective.
Fig.~\ref{fig:convergence_tau5_semilogy} shows that most of the objective reduction occurs during the early iterations, while later iterations mainly refine the solution. The convergence speed also depends strongly on the connectivity structure: \(G=64\) reaches a negligible remaining gap fastest, \(G=8\) exhibits intermediate behavior, and \(G=1\) converges most gradually. This trend is consistent with the larger block dimension and stronger coupling associated with more connected BD-RIS architectures. Together with the NMSE results above, \(G=8\) provides an attractive compromise, retaining most of the fully connected optimization gain while exhibiting substantially faster numerical convergence.
%
%
%
\section{Conclusion}
We studied the uplink of a multi-BD-RIS-assisted UC cell-free massive MIMO system under imperfect CSI and pilot reuse. We derived the aggregate channel MMSE statistics and closed-form uplink SINR, and developed a Riemannian optimization method to minimize the long-term sum-NMSE, subject to the symmetry and unitarity constraints of the group-connected BD-RIS architecture. Numerical results showed that group connectivity retains most of the fully connected gain
and that the optimization becomes increasingly effective as the direct AP--user links weaken. Increasing the BD-RIS size further improves NMSE, but this does not necessarily result in higher uplink SE, highlighting the need for future designs that better align channel estimation quality with data-detection performance.
\bibliographystyle{IEEEbib}
\bibliography{References}

\end{document}